\documentclass[twocolumn,aps,prl,superscriptaddress,amsmath,amssymb]{revtex4-2}
\usepackage{amssymb}
\usepackage{amsmath}
\usepackage{amsfonts}
\usepackage{graphicx}
\usepackage{color}
\usepackage{bm}
\usepackage[colorlinks,linkcolor=blue,anchorcolor=blue,citecolor=blue,urlcolor=blue]{hyperref}
\usepackage[T1]{fontenc}    % 
\usepackage{lmodern}
\usepackage{fix-cm}
\usepackage[utf8]{inputenc} % 
\begin{document}

\title{Evidence of Crystal-Field-Mediated Anisotropic Orbital Rashba Effect in Epitaxial Ru/FM Heterostructure}

\affiliation{State Key Laboratory of Semiconductor Physics and Chip Technologies, Institute of Semiconductors, Chinese Academy of Sciences, Beijing 100083, China}
\affiliation{Center of Materials Science and Optoelectronics Engineering, University of Chinese Academy of Sciences, Beijing 100049, China}
\affiliation{School of Integrated Circuit Science and Engineering, Beihang University, Beijing 100191, China}
\affiliation{State Key Laboratory of Spintronics, Hangzhou International Innovation Institute, Beihang University, Hangzhou 311115, China}

\author{Yu Zhang}
\affiliation{State Key Laboratory of Semiconductor Physics and Chip Technologies, Institute of Semiconductors, Chinese Academy of Sciences, Beijing 100083, China}
\affiliation{Center of Materials Science and Optoelectronics Engineering, University of Chinese Academy of Sciences, Beijing 100049, China}

\author{Wenhao Niu}
\affiliation{School of Integrated Circuit Science and Engineering, Beihang University, Beijing 100191, China}

\author{Yumin Yang}
\affiliation{State Key Laboratory of Semiconductor Physics and Chip Technologies, Institute of Semiconductors, Chinese Academy of Sciences, Beijing 100083, China}
\affiliation{Center of Materials Science and Optoelectronics Engineering, University of Chinese Academy of Sciences, Beijing 100049, China}

\author{Wenqi Xu}
\affiliation{State Key Laboratory of Semiconductor Physics and Chip Technologies, Institute of Semiconductors, Chinese Academy of Sciences, Beijing 100083, China}
\affiliation{Center of Materials Science and Optoelectronics Engineering, University of Chinese Academy of Sciences, Beijing 100049, China}

\author{Gengchen Meng}
\affiliation{State Key Laboratory of Semiconductor Physics and Chip Technologies, Institute of Semiconductors, Chinese Academy of Sciences, Beijing 100083, China}
\affiliation{Center of Materials Science and Optoelectronics Engineering, University of Chinese Academy of Sciences, Beijing 100049, China}

\author{Hailong Wang}
\affiliation{State Key Laboratory of Semiconductor Physics and Chip Technologies, Institute of Semiconductors, Chinese Academy of Sciences, Beijing 100083, China}
\affiliation{Center of Materials Science and Optoelectronics Engineering, University of Chinese Academy of Sciences, Beijing 100049, China}

\author{Na Lei}
\email[Corresponding author: ]{na.lei@buaa.edu.cn}
\affiliation{School of Integrated Circuit Science and Engineering, Beihang University, Beijing 100191, China}
\affiliation{State Key Laboratory of Spintronics, Hangzhou International Innovation Institute, Beihang University, Hangzhou 311115, China}

\author{Dahai Wei}
\email[Corresponding author: ]{dhwei@semi.ac.cn}
\affiliation{State Key Laboratory of Semiconductor Physics and Chip Technologies, Institute of Semiconductors, Chinese Academy of Sciences, Beijing 100083, China}
\affiliation{Center of Materials Science and Optoelectronics Engineering, University of Chinese Academy of Sciences, Beijing 100049, China}

\begin{abstract}
Electrical generation of orbital angular momentum provides a promising route to current-induced torques, 
yet effective control of orbital Rashba textures still remains challenging, particularly because the role of the interfacial crystal field remains largely unexplored. 
Here, we report experimental evidence for a crystal-field-mediated interfacial anisotropic orbital Rashba effect (AORE) in epitaxial Ru/ferromagnet heterostructures. 
Total orbital torque was disentangled into an isotropic bulk contribution and an in-plane anisotropic interfacial contribution. 
The latter was strongly suppressed by degrading the crystallinity of either constituent and by inserting a Cu spacer, highlighting the essential roles of coherent interfacial orbital hybridization and direct Ru/ferromagnet contact. 
These results identify interfacial crystal-field coherency as a key ingredient in manipulating orbital Rashba textures and establish a route toward engineering the symmetry and directionality of orbital torques.
\end{abstract}

\maketitle
%The interconversion among spin angular momentum (SAM), orbital angular momentum (OAM), and charge constitutes the central framework of spin-orbitronics[1]. 
The interconversion among angular momentum and charge constitutes the central framework of spin-orbitronics \cite{Main_Yesdas_APL_2026}.
%Over the past decade, spin transport has been extensively investigated, leading to a relatively unified physical picture, while the recently emerging field of orbitronics remains at an exploratory stage[2]. 
Over the past decade, spin transport has been extensively investigated, while the recently emerging field of orbitronics remains at an exploratory stage\cite{Main_Fukami_NatPhys_2025}. 
%To date, various mechanisms for the generation and transport of orbital currents in different material systems have been theoretically predicted and experimentally demonstrated[3-9], while magnetization switching driven by OAM has also been observed and discussed[10,11]. 
Various mechanisms for the generation of orbital currents have been theoretically predicted and experimentally demonstrated\cite{Main_Go_PRL_2023,Main_Go_PRL_2018,Main_Ding_PRL_2024,Main_Wang_PRB_2025,Main_Wang_NatMater_2025}, 
while magnetization switching driven by orbital angular momentum (OAM) has also been discussed\cite{Main_Shin_AFM_2025,Main_Zheng_NatComm_2024}. 
Despite these advances, the effective manipulation on OAM generation remains far less developed than that of its spin counterpart\cite{Main_Bihlmayer_NatRevPhys_2022}. 
Establishing an efficient means to manipulate OAM is therefore essential for further exploiting orbital degrees of freedom in spin-orbitronic devices.

The challenge in controlling orbital currents originates from the microscopic nature of OAM, which differs fundamentally from spin angular momentum.
In particular, OAM is intrinsically sensitive to the local crystal field, which strongly influences its magnitude and dynamics\cite{Main_Atencia_APX_2024}.
Although the equilibrium OAM was quenched by the crystal field, an external electric field can still excite a finite orbital current\cite{Main_Go_PRL_2018}. 
This dual role suggests that the crystal field should not merely be regarded as a source of orbital quenching, but can instead provide an effective degree of freedom for manipulating nonequilibrium orbital responses.
Through modifying the local orbital occupation and hybridization between different orbitals, local crystal field may reshape the orbital states and interorbital hopping processes underlying OAM generation and transport.
Previous studies have shown that the crystal field can substantially affect the dephasing, relaxation, and transport lifetime of OAM , motivating its use as a route toward active control of orbital transport\cite{Main_Kang_NatPhys_2026}.

As a region where the crystal field varies sharply in real space, the nonmagnetic/ferromagnetic (NM/FM) interface may play a crucial role in orbital-current generation and transport,
where broken translational symmetry and interfacial bonding give rise to orbital hybridization distinct from that of the constituent bulk materials\cite{Main_Go_PRR_2020,Main_Johansson_JPCM_2024,Main_Okano_PRL_2019,Main_Pezo_PRB_2025,Main_Yan_PRB_2026}. 
Early studies demonstrated that interfacial orbital textures do exist even in the absence of significant spin-orbit coupling\cite{Main_Park_PRL_2011,Main_Go_PRB_2021,Main_Go_SciRep_2017},
and has been associated with current-induced orbital accumulation\cite{Main_Nikolaev_NanoLett_2024}, orbital torque\cite{Main_Huang_NanoLett_2023} and magnetoresistance\cite{Main_Ding_PRL_2022,Main_Schmitt_Science_2026}.
Since these orbital responses are governed by the local orbital configuration and interorbital hopping, the interfacial crystal field provides a natural means of directly modifying the microscopic processes responsible for interfacial OAM generation and transport. 
Nevertheless, previous studies have mainly focused on the relationship between crystal symmetry, OAM generation, and dynamical orbital responses\cite{Main_Hayashi_NanoLett_2025,Main_Gao_NatPhys_2024}, whereas the role of the interfacial crystal field in controlling orbital transport remains largely unexplored.

In this work, Ru/FM (FM = Ni, Co, CoFeB) heterostructures with different degrees of structural order were constructed, thereby forming Ru/FM interfaces with different crystal-field coherency (CFC). 
Through measuring orbital torques along different in-plane directions, an interfacial but anisotropic contribution was observed which strongly depends on the crystallinity of the Ru/FM interface. 
Either disrupting direct Ru/FM contact or reducing the Ru or FM layer from a single-crystalline to an amorphous state would eliminates such anisotropy torque. 
Based on our experimental results, we propose a crystal-field-mediated anisotropic orbital Rashba effect (AORE), in which a coherent crystal field reshapes the orbital textures and imprints crystallographic anisotropy on the resulting OAM response. 
These results establish interfacial crystal-field engineering as a new strategy for tailoring orbital textures and controlling orbital transport.\\

\emph{Observation of Crystalline Dependent Orbital Torque.}--
A series of Ru(t)/FM heterostructures were fabricated, with FM corresponding to Ni, Co and Co$_{40}$Fe$_{40}$B$_{20}$ (CoFeB) and the thickness of Ru were 2.5-10 nanometers. 
Ru layer was epitaxially deposited on $c$-Al$_{2}$O$_{3}$ substrate through magnetron sputtering. 
Detailed sample preparation procedure were shown in Supplementary Note 1\cite{Supp_Mat}. 
To examine the crystalline structure of our sample, X-Ray diffraction (XRD) and reflection high-energy electron diffraction (RHEED) was performed, which were shown in Supplementary Note 2\cite{Supp_Mat}. 
Structural characterization confirms the growth of a high-quality epitaxial Ru with an atomically flat surface\cite{Main_Milosevic_JAP_2019}. 
XRD results further show that room-temperature-grown Ni on Ru(0001) is exclusively (111) oriented\cite{Main_Higuchi_JJAP_2011}, while Co contains both hexagonal (0002)- and cubic (111)-oriented components. 
CoFeB remains completely amorphous. 
%Thus, three heterostructures with progressively reduced structural coherence across the Ru/FM interface was constructed.

The sample geometry and Harmonic Hall Voltage (HHV) measurement setup were shown in Fig.\ref{fig 1}(a). 
Basic electrical and magnetic properties were shown in Supplementary Note 3 and 4\cite{Supp_Mat}. 
These results confirm that the electric conductivity of Ru lies well within the good-metal regime and the magnetic layer shows a robust in-plane magnetic anisotropy (IMA).  
HHV measurement was carried out to determine the torque efficiency. 
Considering only $\sigma_{y}$ component of OAM polarization, the second-order HHV $R_{xy} ^{2\omega}$ could be expressed as\cite{Main_Hayashi_PRB_2014,Main_Liu_AdvMater_2024}:

\begin{equation}
	\begin{aligned}
	&R_{xy} ^{2\omega} = (R_{AHE} \frac{H_{DL}}{H_{xy}+H_{k}}+R_{ANE})cos\varphi\\
	&+2R_{PHE}\frac{H_{FL+Oe}}{H_{xy}}(2cos^{3}\varphi-cos\varphi)\\
	\end{aligned}\label{eq:steady}
\end{equation}

\begin{figure}
	\centering
	\includegraphics[width=1\columnwidth]{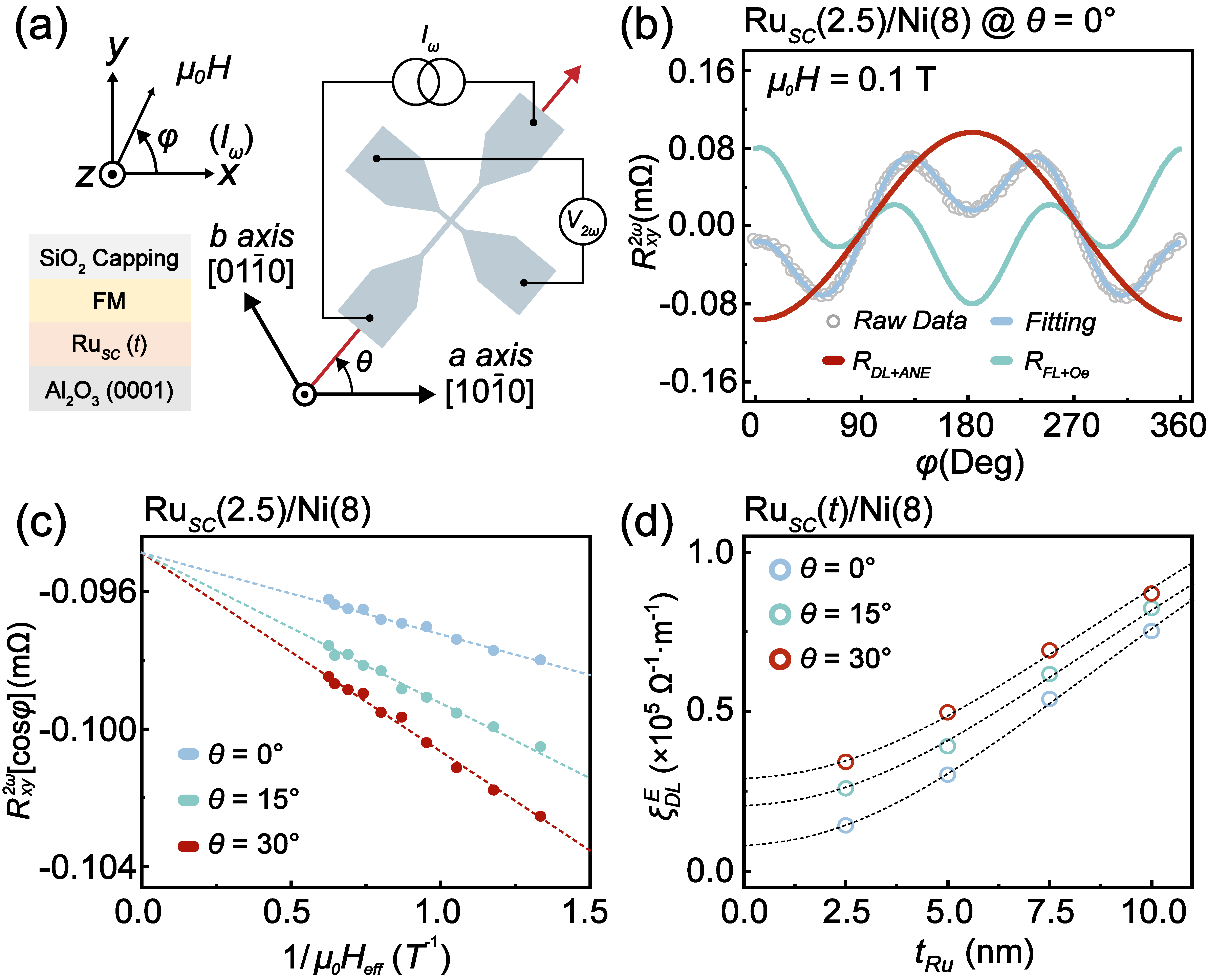}
	\caption{HHV results for Ru/Ni series samples. 
	(a) Schematic illustration of HHV measurement and sample geometry. 
	(b) $R_{xy}^{2\omega}$ results for Ru$_{SC}$(2.5)/Ni(8) at $\mu_{0}H = 0.1 T$. 
	The applied electric field was along $[10\bar10]$. 
	(c) Extracted $R_{xy}^{2\omega}$ as a function of $1/(H_{xy}+H_{k})$ in Ru$_{SC}$(2.5)/Ni(8) at $\theta$ = 0°, 15° and 30°, respectively. 
	(d) $\xi_{DL}^{E}$ as a function of $t_{Ru}$, measured at $\theta$ = 0°, 15° and 30°, respectively. Black curves were fits to the diffusive transport model.
	}
	\label{fig 1} 
\end{figure}

Where $H_{k}$ was the in-plane anisotropy field, $H_{xy}$ was the applied magnetic field. 
$R_{AHE}$, $R_{ANE}$ and $R_{PHE}$ were the anomalous Hall, anomalous Nernst and planar Hall resistances, respectively. 
$H_{DL}$ was the effective Damping-like (DL) field, $H_{FL+Oe}$ was the sum of Field-like (FL) effective field and Oersted field. 

To examine the crystalline-dependent orbital torque, we patterned Hall bar along different in-plane directions.  
Here we define $[10\bar10]$ as a axis and $[01\bar10]$  as b axis. 
The angle between the applied electric field and $a$ axis was defined as $\theta$. 
At each $\theta$, the electric-field-normalized DL torque efficiency was determined through:

\begin{equation}
	\begin{aligned}
	&\xi_{DL}^{E}(\theta)=\frac{2e}{\hslash}\frac{\mu_{0}M_{s}t_{FM}H_{DL}(\theta)}{E}\\
	\end{aligned}\label{eq:steady}
\end{equation}

Where $e$ was the elementary charge, $\hbar$ was the reduced Planck constant, $\mu_{0}$ was the vacuum permeability, $M_{s}$ was the saturation magnetization, $t_{FM}$ was the thickness of ferromagnetic layer, $E$ was the electric field strength. 
We first examine the $Ru_{SC}(t)/Ni(8)$ series sample. 
$R_{xy}^{2\omega}$ raw data for Ru$_{SC}$(2.5)/Ni(8) was shown in Fig.\ref{fig 1}(b). 
The extracted cos$\varphi$ term as a function of $1/\mu_{0}(H_{xy}+H_{k})$ was shown in Fig.\ref{fig 1}(c), which exhibits a significant dependence on crystalline direction $\theta$. 
$\xi_{DL}^{E}$ as a function of $t_{Ru}$ was shown in Fig.\ref{fig 1}(d). 
PHE measurement at various magnetic field were shown in Supplementary Note 5, which exhibit a clear sinusoidal dependence\cite{Supp_Mat}. 
We note that magnetic relaxation or angular-momentum absorption process within the FM has previously been reported and likely to generate orientation-dependent torques through a magnetization-direction-dependent damping\cite{Main_Chen_NatPhys_2018,Main_Baker_PRL_2016}. 
In our measurements, however, the anisotropic torque was governed by Ru crystallographic axes, 
while the magnetization undergoes coherent in-plane rotation, making a dominant FM-relaxation origin unlikely. 
A detailed discussion on self-torque was shown in Supplementary Note 6\cite{Supp_Mat}, the results demonstrate a sizable self-torque in Ni, in agreement with previous studies\cite{Main_Liu_NatComm_2025}, and further exclude nonuniform magnetic textures in the Ni layer as the origin of the observed anisotropic torque.

\emph{Separation of Isotropic and Anisotropic Torque Contribution.}--
To further examine the crystalline-dependent torque, additional torque measurements were carried out in $Ru/Ni(8)$, as shwon in Fig.\ref{fig 2}(a). 
Either introducing a 1-nm-thick Cu insertion layer or use amorphous Ru layer, the crystalline-dependent torque completely vanished, as shown in the blue and gray curve of Fig.\ref{fig 2}(a). 
Such results indicate that the observed anisotropic torque strongly dependent on the crystalline of Ru and Ru/FM interface. 
Thus, total torque could be decomposed into the following parts:

\begin{figure}
	\centering
	\includegraphics[width=1\columnwidth]{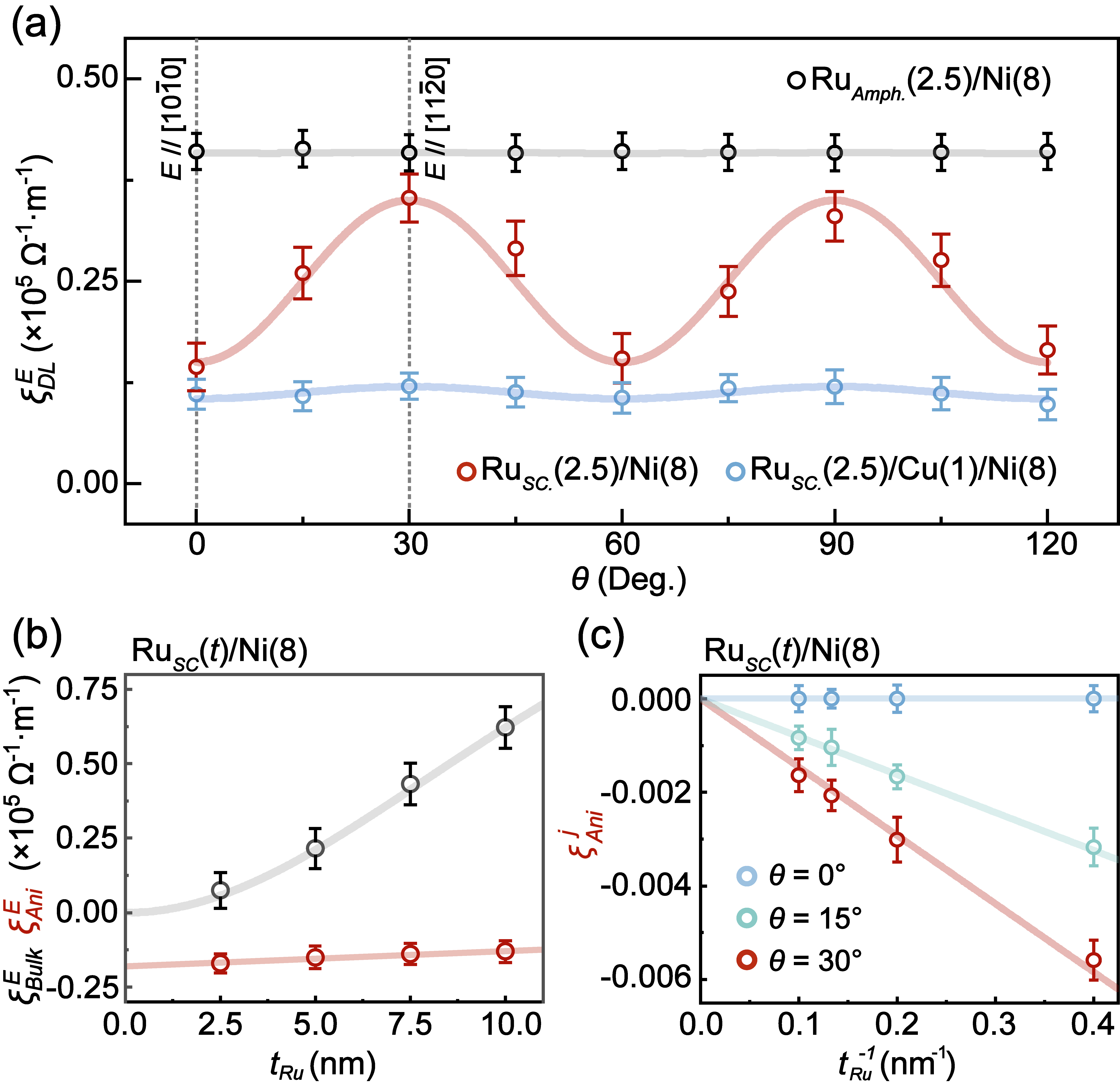}
	\caption{Crystalline dependence of orbital torque. 
	(a) $\xi_{DL}^{E}$ for Ru$_{SC}$(2.5)/Ni(8), Ru$_{SC}$(2.5)/Cu(1)/Ni(8) and Ru$_{Amph}$(2.5)/Ni(8) samples at different in-plane crystalline direction. 
	(b) Extracted $\xi_{Bulk}^{E}$ and $\xi_{Ani}^{E}$ terms as a function of $t_{Ru}$. 
	(c) Extracted $\xi_{Ani}^{j}$ as a function of $t_{Ru}^{-1}$ at $\theta$ = 0°, 15° and 30°.
	}
	\label{fig 2} 
\end{figure}

\begin{equation}
	\begin{aligned}
	&\xi_{Total}^{E}=\xi_{Bulk}^{E}+\xi_{Ani}^{E}+\xi_{FM}^{E}\\
	\end{aligned}\label{eq:steady}
\end{equation}

where $\xi_{Bulk}^{E}$ was the isotropic bulk contribution of Ru, $\xi_{Ani}^{E}$ was the crystalline-dependent anisotropic term and $\xi_{FM}^{E}$ was the self-torque of FM layer. 
Here we define $\xi_{Ani}^{E}=\xi_{Ani}^{[10\bar10]}-\xi_{Ani}^{[11\bar20]}$,which was a negative value. 
We note that either introducing Cu layer or using amorphous Ru as a control would inevitably introduce additional interfacial torque contributions, modify the interfacial transparency and substantially change the orbital Hall conductivity (OHC) of Ru. 
A more appropriate approach is to measure the self-torque of the FM layer and extrapolate the bulk contribution to $t_{Ru}=0$. 
A detailed discussion on torque extraction was shown in Supplementary Note 6\cite{Supp_Mat}. 
The extracted $\xi_{Bulk}^{E}$ and $\xi_{Ani}^{E}$ was shown in Fig.\ref{fig 2}(b). 
The two contributions have opposite signs. 
$\xi_{Bulk}^{E}$ increases systematically with Ru thickness, consistent with a bulk origin, whereas $\xi_{Ani}^{E}$ exhibits little thickness dependence, indicates an interfacial origin. 
A slightly decreases of $\xi_{Ani}^{E}$ as Ru thickness increases could be attributed to the gradual deterioration of the Ru interface quality.
Detailed electrical conductance and thickness-dependent orbital torque for amorphous Ru was shown in Supplementary Note 7\cite{Supp_Mat}. 
The resistivity of amorphous Ru was significantly higher, whereas the overall orbital torque was smaller. 
This may result from the resistivity normalization applied in the torque analysis.

As expected, isotropic $\xi_{Bulk}^{E}$ obeys the diffusive transport model and follows the relation of\cite{Main_Nguyen_PRL_2016}:

\begin{equation}
	\begin{aligned}
	&\xi_{Bulk}^{E}=\frac{2e}{\hbar}\sigma_{OH}[1-sech(\frac{t_{Ru}}{\lambda_{Ru}})](1+\frac{tanh(t_{Ru}/\lambda_{Ru})}{2\lambda_{Ru}\rho_{Ru}G_{r}})\\
	\end{aligned}\label{eq:steady}
\end{equation}

where $G_{r}$ was the real part of mixing conductance, $\lambda_{Ru}$ was the diffusion length in Ru. 
The calculated $\lambda_{Ru}$ was 9.5$\pm$1.8 nm, which was shorter than previous reported results in amorphous Ru/FM\cite{Main_Xu_APL_2026}. 
This could be attributed to the enhanced crystal field in single crystal Ru layer which affects the orbital relaxation time and orbital conductivity\cite{Main_Kang_NatPhys_2026}. 
However, we note that due to the limited number of data points used for fitting, the obtained value of $\lambda_{Ru}$ in our results should be regarded as an approximate estimate rather than a precise value.
The current-density-normalized anisotropic torque term $\xi_{Ani}^{j}$ was plotted against $1/t_{Ru}$, as shown in Fig.\ref{fig 2}(c). 
The linear dependence of $1/t_{Ru}$ further demonstrated that $\xi_{Ani}$ was originated from the Ru/FM interface. 
Similar HHV measurements were also carried out for Ru$_{SC}$(t)/Co(5) and Ru$_{SC}$(t)/CoFeB(8) series samples, which were shown in Supplementary Note 8\cite{Supp_Mat}. 
Fig.\ref{fig 3}(a) compares the $\xi_{Ani}^{E}/\xi_{Bulk}^{E}$ ratio for Ru$_{SC}$(t)/Ni, Ru$_{SC}$(t)/Co, and Ru$_{SC}$(t)/CoFeB.
$\xi_{Ani}^{E}$ was suppressed in Ru$_{SC}$/Co and nearly vanishes in Ru$_{SC}$/CoFeB. 
Such results strongly indicate the dominant role of Ru/FM crystalline.
Analysis of FL torque and Oersted contribution was performed in Supplementary Note 9\cite{Supp_Mat}. 
FL and DL torques exhibit similar angular trends, indicates their common physical origin.\\

\begin{figure*}[t]
	\centering
	\includegraphics[width=0.9\textwidth]{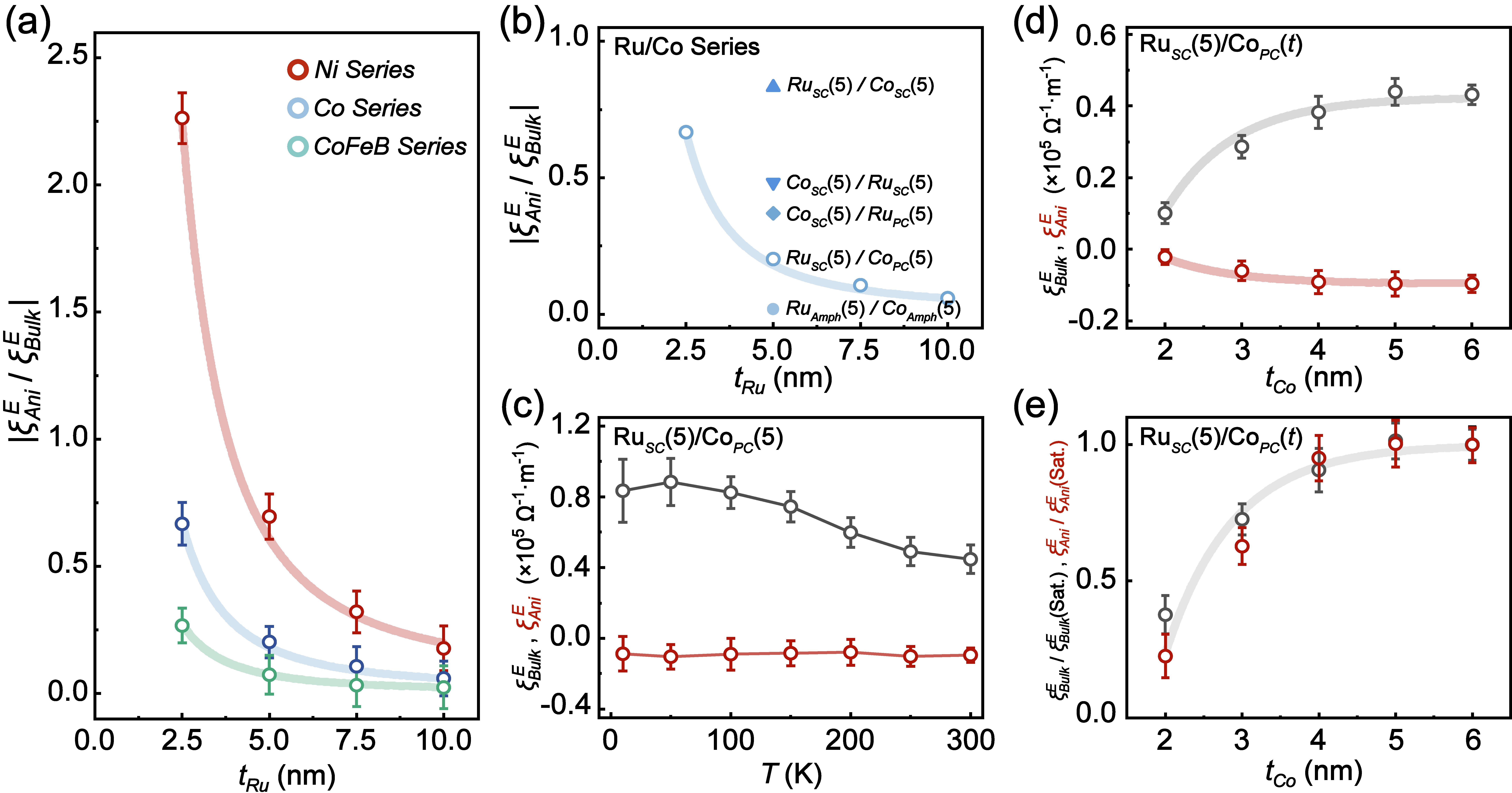}
	\caption{ 
	(a) Comparison of $|\xi_{Ani}^{E}/\xi_{Bulk}^{E}|$ for Ru$_{SC}$(t)/Ni, Ru$_{SC}$(t)/Co and Ru$_{SC}$(t)/CoFeB series samples as a function of t$_{Ru}$. 
	(b) Detailed comparison of $|\xi_{Ani}^{E}/\xi_{Bulk}^{E}|$ results for Ru/Co series sample, with different interfacial crystalline and stacking order. 
	(c) Temperature dependence of $\xi_{Bulk}^{E}$ and $\xi_{Ani}^{E}$ terms in Ru$_{SC}$(5)/Co$_{PC}$(5) sample. 
	(d) Extracted $\xi_{Bulk}^{E}$ and $\xi_{Ani}^{E}$ in Ru$_{SC}$(5)/Co$_{PC}$(t) as a function of t$_{Co}$. 
	(e) Scaling of $\xi_{Bulk}^{E}$ and $\xi_{Ani}^{E}$ data with its values at t$_{Co}$=6 nm. 
	The light gray line was a fit to $\frac{\xi_{(Ani)Bulk}^{E}(t_{Co})}{\xi_{(Ani)Bulk}^{E}(Sat.)}\propto 1-exp(\frac{-t_{Co}}{\lambda_{Co}})$. 
	}
	\label{fig 3} 
\end{figure*}

\emph{Origin of Interfacial Anisotropic Orbital Torque.}--
To systematically examine this issue, Ru/Co heterostructures with different crystallinities and stacking sequences were fabricated. 
A detailed growth procedure, structural and magnetization characterization were shown in Supplementary Note 1 and 10\cite{Supp_Mat}. 
Co was chosen as the FM layer for these measurements was because the magnetic properties of Ni are highly sensitive to interface conditions, thickness and temperature \cite{Main_Farle_JMMM_1997,Main_Zhang_JPCM_2024}. 
As shown in Fig.\ref{fig 3}(b), $\xi_{Ani}^{E}/\xi_{Bulk}^{E}$ was largest for Ru$_{SC}$/Co$_{SC}$, decreases when either layer becomes polycrystalline, and becomes negligible in amorphous structure
, demonstrating a strong correlation between $\xi_{Ani}^{E}$ and interfacial structural order.

Temperature-dependent results provide further evidence that $\xi_{Ani}^{E}$ and $\xi_{Bulk}^{E}$ share different microscopic origins, as shown in Fig.\ref{fig 3}(c). 
Detailed sample resistivity, R$_{AHE}$ and $M_{s}$ at low temperature were shown in Supplementary Note 11\cite{Supp_Mat}. 
For Ru$_{SC}$(5)/Co$_{PC}$(5), $\xi_{Bulk}^{E}$ increases substantially upon cooling, consistent with previous studies \cite{Main_Xu_APL_2026}, while $\xi_{Ani}^{E}$ remains basically unchanged. 
Orbital torque for various FM-thickness was carried out in Ru$_{SC}$(5)/Co$_{PC}$(t). 
As shown in Fig.\ref{fig 3}(d), the magnitudes of both contribution increase with Co thickness and gradually comes to saturation. 
When normalized to their saturated values, both terms collapse onto the same scaling curve, described by orbital diffusion, as shown in Fig.\ref{fig 3}(e). 
This common scaling behavior suggests that, despite their different generation mechanisms, they both originated from OAM.

A further question is whether the observed anisotropic torque arises from an orbital or a spin channel. 
Anisotropic spin Hall conductivities in hcp metals have been predicted from first principles \cite{Main_Freimuth_PRL_2010}
, and an anisotropic interfacial spin Rashba response is therefore an a priori alternative explanation that must be examined. 
Three observations favor the orbital interpretation: 
(i) the magnitude of the measured torque, 
(ii) the long length scale $\lambda_{Ru}$ ~ 9.5 nm extracted below, 
and (iii) the temperature dependence of the two torque components are all consistent with previously reported orbital channels in Ru\cite{Main_Xu_APL_2026}, whereas they are difficult to reconcile with spin-channel estimates. 
We emphasize that this discrimination rests on the mutual consistency of several indirect lines of evidence rather than on an absolute exclusion of a spin contribution.

\begin{figure*}[t]
	\centering
	\includegraphics[width=0.8\textwidth]{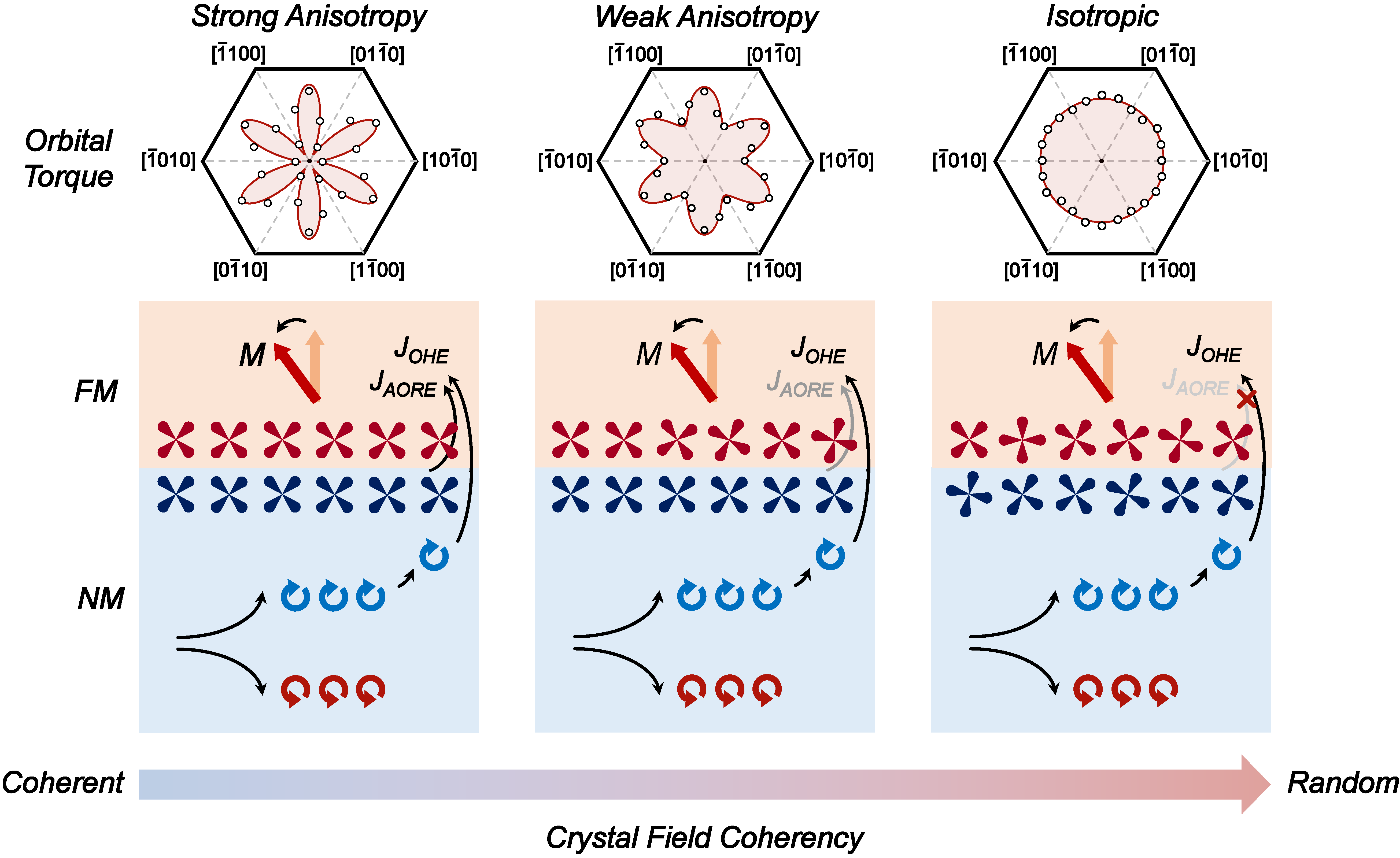}
	\caption{ 
	Schematic illustration of crystal-field-mediated AORE.
	A coherent crystal field modulates the orbital textures and further gives anisotropic ORE response. 
	In the highly coherent limit, CFC modulation produces a pronounced anisotropic orbital response (left panel). 
	As CFC was progressively reduced, such modulation was weakened and the lattice-locked anisotropic component is strongly suppressed (middle panel). 
	In the fully disordered limit, the crystallographic modulation becomes negligible, leaving a predominantly isotropic response (right panel).
	}
	\label{fig 4} 
\end{figure*}

Based on our experimental results, we propose the mechanism of crystal-field-mediated anisotropic orbital Rashba effect (AORE). 
The extended Ru 4$d$ orbitals can directly hybridize with the FM 3$d$ orbitals. 
Interfacial inversion-symmetry breaking makes hopping along the $\pm$$z$ directions inequivalent, thereby enabling ORE. 
We describe the structural dependence as crystal field coherency (CFC), which refers to the spatial correlation of the orientation and symmetry of local crystal-field\cite{Main_Coey_PRL_1976,Main_Uldry_PRB_2012}. 
In a single crystal, common crystallographic axes could preserve this correlation. 
In a polycrystal, it is retained primarily within individual grains and reduced, whereas in an amorphous structure, such coherence was largely lost. 
To elucidate the role of CFC, multiorbital tight-binding calculations were performed for Ru/Co and Ru/Cu/Co, incorporating CFC, spin-orbit coupling and Slater-Koster-type interlayer hybridization 
(as detailed in Supplementary Note 12\cite{Supp_Mat}). 
Structural disorder was induced by varying the CFC of the Co crystal field, while Cu insertion disrupts the direct Ru-Co orbital hybridization. 
The calculated anisotropic ORE response exhibits a clear angular dependence and was strongly suppressed by either reduced CFC or Cu insertion, consistent with the experimental observations. 
We note that the loss of CFC does not eliminate ORE respond itself. Instead, CFC modulates the orbital textures and gives the resulting OAM a preferred crystallographic direction. 
By contrast, OAM generated in bulk Ru was governed by the intrinsic OHC and extrinsic skew scattering. 
Owing to its relatively long mean free path and the symmetry-protected character of the bulk response, it was less sensitive to interfacial CFC and therefore unlikely to be effectively reconstructed by interfacial crystal field.
The total orbital current can be therefore decomposed as $j=j_{OHE}+j_{AORE}$, as shown in Fig.\ref{fig 4}. 
The latter requires a well-defined orientational relationship and CFC among the local crystal-field. 
In the highly coherent limit, $j_{AORE}$ retains pronounced crystallographic selectivity, leading to strongly anisotropic orbital accumulation. 
As the CFC progressively reduced, AORE contribution was suppressed and gradually degenerated into conventional isotropic Rashba-type torque in the fully disordered limit. 
The resulting orbital accumulation and orbital torque in the ferromagnet therefore become isotropic. 
We note that in-plane anisotropic torques have also been reported in CuPt/CoPt\cite{Main_Zhao_AdvMater_2023,Main_Zhao_NatComm_2025,Main_Liu_NatNanotech_2021}, CoPt single layer\cite{Main_Li_AdvFuncMater_2024,Main_Liu_NatComm_2022}, FeSn/Py\cite{Main_Gupta_AdvFuncMater_2025}. 
In $L1_{1}$-CuPt/CoPt and CoPt single layer, the anisotropic torque was mainly associated with a symmetry-allowed $3m$ torque arising from lowered crystal symmetry. 
In FeSn/Py, the six-fold anisotropic torque was attributed to the band structure and Berry curvature of Kagome FeSn. 
More recently, similar results was observed in epitaxial CuO/Ni and was attributed to the direct generation of crystal-symmetry-dependent ORE \cite{Main_Xiao_NatComm_2026}.
Compared with these results, our AORE originates from a crystal-field-modulated interfacial orbital Rashba response, which was fundamentally different.

\emph{Conclusion.}--
In conclusion, we give strong evidence on the existence of anisotropic orbital Rashba effect (AORE) in epitaxial Ru/FM structure. 
AORE exhibits Ru-thickness independence, and was strongly suppressed by Cu insertion or by the loss of long-range crystalline order. 
Combined with the multiorbital tight-binding calculations, the results support a crystal-field-mediated AORE arising from Ru 4$d$-FM 3$d$ orbital hybridization. 
A coherently ordered crystal field produces pronounced crystallographic re-shaping in the nonequilibrium orbital textures. 
Progressive structural disorder suppresses the anisotropic component, while leaving the comparatively isotropic bulk contribution dominant. 
Our work therefore extends crystal-field engineering from equilibrium orbital-level control to nonequilibrium OAM generation and identifies interfacial crystal-field coherency for designing orbitronic devices.

\emph{Acknowledgments}--
D.W. acknowledges the support by National Natural Science Foundation of China (No.12674154).

\end{document}